\documentclass[preprint,showpacs,preprintnumbers,amsmath,amssymb]{revtex4}

\usepackage{revsymb}
\usepackage[english]{babel}
\usepackage{graphicx, color, epsfig}
\usepackage{enumerate}
\usepackage{latexsym}
\usepackage{fancyhdr}
\usepackage{multirow}
\usepackage{dcolumn}
\usepackage{bm}

\newcommand{\ds}{\displaystyle}
\newcommand{\vs}{\vspace}
\newcommand{\nt}{\nonumber}

\newcommand{\C}{\textrm{\scriptsize C}}
\newcommand{\dd}{\mathrm{d}}

\newcommand{\e}{\textrm{\scriptsize e}}
\newcommand{\ee}{\mathrm{e}}
\newcommand{\F}{\mathrm{F}}
\newcommand{\g}{\textrm{\scriptsize g}}
\newcommand{\G}{\textrm{\scriptsize G}}
\newcommand{\J}{\mathrm{J}}
\newcommand{\h}{\textrm{\scriptsize h}}
\newcommand{\kk}{\mathbf{k}}
\newcommand{\p}{\mathbf{p}}
\newcommand{\rr}{\mathbf{r}}

\newcommand{\Y}{\textrm{Y}}

\newcommand{\sigmag}{\textrm{\mathversion{bold}$\sigma$}}
\newcommand{\zerog}{\textrm{\mathversion{bold}$0$}}

\newcommand{\ldc}{[\![}
\newcommand{\rdc}{]\!]}
\newcommand{\pr}{\partial}

\renewcommand{\stackrel}[2]{\!\!
  \raisebox{-0.1cm}{$\begin{array}{c}
#1
\\[-0.25cm]
\textrm{\tiny $#2$}
  \end{array}$}\!\!}

\newcommand{\lnd}{~\!\!\!\!&}
\newcommand{\rnd}{&\!\!\!\!~}

\newcommand{\Si}{\textrm{Si}}

\begin{document}
\title{New Theoretical Approach to Quantum Size Effects of Interactive
Electron-hole in Spherical Semiconductor Quantum Dots}
\author{B. Billaud}
\altaffiliation[Also at ]{LPTM, Universit\'e de Cergy-Pontoise}
\email{baptiste.billaud@u-cergy.fr}
\author{M. Picco}
\affiliation{Laboratoire de Physique Th\'eorique et Hautes Energies (LPTHE),
\\
CNRS UMR 7589, Universit\'e Pierre et Marie Curie (Paris VI),
\\
4, place Jussieu, 75252 Paris Cedex 05, France.}
\author{T.-T. Truong}
\affiliation{Laboratoire de Physique Th\'eorique et Mod\'elisation (LPTM),
\\
CNRS UMR 8089, Universit\'e de Cergy-Pontoise,
\\
2, avenue Adolphe Chauvin, 95302 Cergy-Pontoise Cedex, France.}

\date{\today}

\begin{abstract}
The issue of quantum size effects of interactive electron-hole
systems in spherical semiconductor quantum dots is put to question.
A sharper theoretical approach is suggested based on a new
pseudo-potential method. In this new setting, analytical
computations can be performed in most intermediate steps
lending stronger support to the adopted physical assumptions. The resulting numerical
values for physical quantities are found to be much closer to the
experimental values than those existing so far in the literature.
\end{abstract}


%\submitto{}

\maketitle
\section{Introduction}
For about three decades, nanostructures --- produced by many
techniques such as etching, local inter-diffusion, particle
suspension in dielectric media, or by self-assembly in matrices of a
host material --- are known as confined electronic systems because
they are designed to restrict particle motion to a space region with
a restricted number of dimensions. Depending on their
dimensionality, these structures are called quantum dots (0D),
quantum wires (1D) or quantum wells (2D). The  principal property of
interest resides in their adjustable quantized energy spectrum. The
corresponding wave functions are localized within the confined
region, but may extend over many periods of its boundary crystal
lattice. A quantum dot (QD) is therefore a giant artificial atom
which enjoys prospects for an increasing range of future
applications: {\it e.g.} as a semiconductor laser
\cite{Kirkstaedter_1994}, as {\it qubits} for quantum information
processing, as single-electron transistors in electronics, as
artificial {\it fluorophores} for intra-operative detection of
tumors, biological imaging or cell studies, {\it etc.}

QD properties depend on their electronic structure. In fact, they
contain a finite number of elementary charge carriers (from a few to
a hundred), which may be conduction band electrons, valence band
holes or excitons of the host substrate. There exists a vast number
of aspects related to their own interactions as well as with the
ambient electromagnetic field. Correlation effects
\cite{Bondarenko_2005, Lozovik_2003}, interactions with boundary
electromagnetic modes \cite{Rotkin_2000}, phonon
interactions \cite{Fomin_1998}, magnetic field effects
\cite{Asari}, just to cite only a few topics, are
currently vigorously investigated.

In the early eighties, quantum size effects (QSE) occurred
experimentally in spherical semiconductor QDs in the exploration of
the optical properties of semiconductor microcrystals in an
insulating matrix \cite{Ekimov_1980, Golubkov_1981}. This phenomena
has been observed in a large range of other confined structures:
{\it e.g.} in quantum ribbons or in quantum disks \cite{Kash_1986},
in quantum wires \cite{Temklin_1987} and, indeed, in quantum wells
\cite{Vojak_1980}. It emerges in a widening of the semiconductor
optical band gap. This is due to the increase of confinement energy
for decreasing QD size. The valence band drops and the conduction
band rises. Both effects constitute the leading contribution to the
characteristic blue-shift in the semiconductors optical spectra of
such strongly quantum-confined systems \cite{Efros_1982, Brus}.
So, the confinement of oppositely charged carriers does
have significant effects on the electron-hole Coulomb interaction
and, therefore, on exciton formation.

In order to apprehend correctly the origin of these quantum size
effects, a first theoretical attempt to describe electronic
properties dependent semiconductor QDs has been elaborated upon a
{\it particle-in-a-sphere} model in the effective-mass approximation
(EMA) \cite{Efros_1982, Brus,
Kayanuma_1988}, in which both electron and hole behave as
non-interacting electrons and holes, trapped in a spherical infinite
potential well but with different masses. Since 1973, the QD spherical shape has
been used \cite{Baldereschi} and continues to be
very popular over the years \cite{Efros_1982,
Kayanuma_1988, Keller_1995, Sundqvist_2002, Bondarenko_2005}.
Because of the usual assumption of parabolic band structure, their
effective mass is commonly defined through the inverse of the second
derivative of their kinetic energy with respect to their momentum.
Except for a pioneering work \cite{Efros_1982}, the Coulomb
interaction between electron and hole has been included and the
excitonic contribution to the ground state energy has been taking
into account by the Ritz variational principle. Some other authors
have developed their own EMA model based on finite potential wells
and improved agreement with experimental data for a significant
range of QD sizes \cite{Kayanuma_1988, Kayanuma_1990,
Kayanuma_1991}. We should also quote some other variational
calculations, {\it e.g.} \cite{Nair_1987, Thoai_1990, Lo_1991}. In
addition to spherical clusters, the case of cylindrical shaped
microcrystallites has been carefully treated and experimentally
studied \cite{Kayanuma_1990, Potter,
Le_Goff_1992}, as well as the case of quantum wires
\cite{Brown_1986, Degani_1987}. Later on, more sophisticated models
have been conceived. Some of them, called empirical tight-binding
method (ETBM), have considered non-parabolic valence and/or
conduction band(s) \cite{Wang_1990, Nomura_1991}. Others have used a
reformulation of the so-called $\kk\cdot\p$ perturbation theory
including non-parabolic bands \cite{Xia_1989, Sercel_1990,
Vahala_1990} based upon the Baldereschi-Lipari Hamiltonian
\cite{Baldereschi}.

However, despite the existence of numerous theoretical and/or
empirical models, to the best of our knowledge, there exists
actually no simple and comprehensive one, which offers a significant
analytical treatment of the problem. This is the reason why this
paper propose an alternative single EMA model for spherical
semiconductor microcrystals, which allows us to clarify some tricky
points. Moreover, we are able to exhibit the existence of an
effective potential, whose presence significantly changes the QD
ground state at small QD radius. Furthermore, in this formalism, it
finally becomes possible, as an achievement, to analytically
calculate some important quantities, such as {\it e.g.} the function
$\eta$ introduced by Kayanuma in \cite{Kayanuma_1988}, with a good
agreement with experimental results. To this end, we shall introduce
our model and recall some general properties in Sec. {\bf \ref{k}}.
Then, the two next Secs. {\bf \ref{l}} and {\bf \ref{m}} are devoted
to the analysis of the two limiting regimes of semiconductor QDs which we shall explicitely
describe and in which we carry out analytic computations
in some details. Finally, in Sec. {\bf \ref{n}}, the new effective
potential mentioned above is introduced and we show how it leads to
improve numerical results, much closer to experimental data. In the
concluding section, we summarize our main results and indicate possible
future research perspectives.
\section{Quantum Dot model and some general results} \label{k}
Since our aim is to understand the semiconductor optical bands
blue-shift as a main effect, we shall discard spin effects
(electron-hole spin coupling or external applied magnetic field) and
consider non-relativistic spinless electron or hole, trapped in a
confining infinite spherical potential well. There exist other
models with parabolic confinement \cite{Lozovik_2003, Keller_1995}
or parabolic potential superimposed to an infinite potential well
\cite{Sundqvist_2002}, which are used to explain certain
spectroscopic data; but the concept of a QD size is then not so well
defined. Here, we adopt the EMA model previously introduced.
\subsection{Intrinsic limitations}
First, the EMA assumption breaks down when a significant number of
carriers wave functions inside the QD overlap. Thus, our results are
expected to fail for very small nanocrystals. As matter of fact, at typical sizes of
less than a hundred lattice spacings appear, in a semiconductor,
magic numbers of clusters of which only some remain stable: {\it
e.g.} nanocrystalline silicon stay solely coherent as clusters of
$Si_{12}, Si_{33}, Si_{39}$ and $Si_{45}$, if they contain less than
60 silicon atoms \cite{Pan_1994}. When such size is reached, their
band structure should be very deformed and, therefore, cannot satisfy
the parabolic spectrum assumption on which the EMA is based.

Second, by assuming spherical symmetry for the QD, we justify the
splitting of the Schr\"odinger equation into a sum of a radial part
and an angular part. This manipulation appreciably reduces the
computation of the eigenstate wave functions and the energy
eigenvalues. This approximation turns out often to be quite good
since most synthesized nanocrystals possess an {\it aspect ratio}
(defined as the ratio between the longest and shortest axes of the
QD) smaller than 1.1. But, for high aspect ratio microcrystals, this
spherical assumption becomes no longer valid.

Third, the real potential at the boundary of the QD is, of course,
not infinite. Actually, there is a potential step with a standard
height of magnitude from 1 to 3eV \cite{Thoai_1990}. This value is
generally quite large compared to typical electron and hole energies
and, therefore, their tunnel conductivity from the nanocrystal to
its surrounding should be neglected, except for very small QD sizes.
Furthermore, the infinite potential well approximation implies that
the charge carrier behavior inside a QD is totally insensitive to
any externally applied potential or to the surrounding of the
cluster. Although the surrounding effects may be sufficiently small
to be neglected, the presence of a large external potential can
in fact significantly modify the inside behavior of the
microcrystallites. Thus, in order to test the validity of
this approximation, we have to consider charged carriers isolated
from the outside neighboring semiconductor.
\subsection{A free Quantum Dot model}
Let $V(r)$ be the confining potential well, defined in spherical
coordinates as
  $$
V(\rr)=V(r)=\!
    \left\{
      \begin{array}{cccc}
0 & \textrm{if} & 0\leq r\leq R, & \textrm{Region I};
\\
\infty & \textrm{if} & r>R, & \textrm{~Region II}.
      \end{array}
    \right.
  $$
Neglecting, for the moment, the electron-hole Coulomb interaction,
in single parabolic band approximation, the Hamiltonian operator is
(with $\hbar=1$)
  \begin{eqnarray}
H_0\lnd=\rnd H_\e+H_\h+E_g \nt
\\
\lnd=\rnd-\frac{\nabla^2_{\!\e}}{2m_\e^*}-\frac{\nabla^2_{\!\h}}{2m_\h^*}+V(\rr_\e)+V(\rr_\h)+E_\g,
\label{a}
  \end{eqnarray}
where $m_{\e,\h}^*$ and $H_{\e,\h}$ denote the effective mass and
the confined Hamiltonian respectively of the electron and of the
hole, and $E_\g$ is the semiconductor energy band gap. The
semiconductor QD wave function can now be written as a product of its
electronic and hole parts
  $$
\Psi(\rr_\e,\rr_\h)=\psi(\rr_\e)\psi(\rr_\h).
  $$
The orthonormal eigenfunctions $\psi_{lnm}$ are labelled by three quantum numbers:
$l\!\in\!\mathbb{N}$, $n\!\in\!\mathbb{N}^*=\mathbb{N}-\{0\}$ and
$m\!\in\!\ldc-l,l\rdc$.
  \begin{eqnarray*}
\psi_{lnm}(\rr)\lnd=\rnd\psi_{lnm}(r,\theta,\varphi)
\\
\lnd=\rnd\frac{\chi_{[0,R[}(r)}{R\J'_{\nu_l}(k_{ln})}\sqrt{\frac2r}\J_{\nu_l}\!\!\left(\frac{k_{ln}}Rr\!\right)\!\Y^m_l(\theta,\varphi),
  \end{eqnarray*}
where
  \begin{itemize}
\item $\Y^m_l(\theta,\varphi)$ is the spherical harmonic of orbital quantum number $l$ and azimuthal quantum number $m$,
\item $\J_{\nu_l}(x)$ is the Bessel function of the first kind of index $\nu_l\!=\! l+\frac12$ and variable $x$,
\item $\chi_{[0,R[}(r)$ is the characteristic function of radial interval,
\item $\left\{k_{ln}\right\}_{ln}$ are the wave numbers in Region I, defined as the
$n^{\textrm{\scriptsize th}}$ non-zero root of the Bessel function
$\J_{\nu_l}(x)$, thanks to the continuity condition at $r=R$.
  \end{itemize}
The respective energy eigenvalues for electron and hole are
expressed in terms of the same family of wave numbers $\left\{k_{ln}\right\}_{ln}$
  $$
E^{\e,\h}_{ln}=\frac{k_{ln}^2}{2m^*_{\e,\h}R^2}.
  $$
This clearly indicates that the continuum density of states of the
semiconductor bulk should show atomic-like discrete energy levels
with increasing energy separation as the QD radius decreases.
\subsection{Electron-hole pair Quantum Dot model} \label{h}
From now on, we add the Coulomb interaction $V_\C$ between electron
and hole to the Hamiltonian $H_0$, given by Eq. (\ref{a}) and define
the usual spherical shape semiconductor QD Hamiltonian $H$ in the
effective-mass approximation \cite{Brus, Kayanuma_1988, Kayanuma_1990, Kayanuma_1991} as
  \begin{eqnarray}
\!\!\!\!\!H\lnd=\rnd H_0+V_\C(\rr_{\e\h}) \nt
\\
\lnd=\rnd-\frac{\nabla^2_{\!\e}}{2m_\e^*}-\frac{\nabla^2_{\!\h}}{2m_\h^*}+V(\rr_\e)+V(\rr_\h)-\frac{e^2}{\kappa r_{\e\h}}+E_\g,
  \end{eqnarray}
where $\kappa=4\pi\varepsilon$ and $\varepsilon$ is the
semiconductor dielectric constant.

In order to simplify notations, we shall assume that $E_\g=0$.
Deriving an exact analytical solution is arduous because of the
Coulomb potential dependence in the electron-hole relative distance
$r_{\e\h}\!=\!|\rr_{\e\h}|\!=\!|\rr_\e-\rr_\h|$, which explicitly
breaks the spherical symmetry of the system. The common approach to
this problem is to treat differently the interplay of the
Coulomb interaction, which scales as $\propto\!
R^{-1}$, and the quantum confinement, which scales as $\propto
R^{-2}$. To handle these competing contributions, two regimes of
electron-hole pair should be singled out by comparing the order of
magnitude of the QD radius $R$ to the Bohr radius of the bulk {\sc
Mott-Wannier} exciton $\ds a^*=\frac{\kappa}{e^2\mu}$, $\mu$ being
the reduced mass of the exciton. These are
  \begin{itemize}
\item the strong confinement regime, valid for a size $R\leq2a^*$ \cite{Kayanuma_1988},
in which the potential well strongly affects the relative
electron-hole motion, the {\it exciton} states consist then of
uncorrelated electron and hole states;
\item the weak confinement regime, valid for a size $R\geq4a^*$ \cite{Kayanuma_1988},
in which the electron-hole relative motion and its binding energy are {\it quasi} left unchanged.
The exciton could be treated as a confined quasi-particle of total
mass $M=m_\e^*+m_\h^*$ and its center-of-mass motion should be
quantized.
  \end{itemize}

In any case, the Coulomb potential should be treated as a
perturbation of the infinite confinement potential well. So, in order
to evaluate the ground state energy of the exciton state using a
variational procedure, we shall use the following wave function
  \begin{equation}
\phi(\rr_\e,\rr_\h)=\psi_{010}(\rr_\e)\psi_{010}(\rr_\h)\phi_{\textrm{\scriptsize
rel}}(\rr_{\e\h}), \label{e}
  \end{equation}
where $\phi_{\textrm{\scriptsize
rel}}(\rr_{\e\h})=\phi_{\textrm{\scriptsize
rel}}(r_{\e\h})=\ee^{-\frac\sigma2r_{\textrm{\tiny
eh}}}$, $\sigma$ denotes a variational parameter, $r_{\e,\h}=|\rr_{\e,\h}|$, and
  $$
\psi_{010}(\rr_{\e,\h})=\psi_{010}(r_{\e,\h})=-\frac{\chi_{[0,R]}(r_{\e,\h})}{r_{\e,\h}\sqrt{2\pi R}}\sin\!\left(\frac\pi R r_{\e,\h}\right)\!.
  $$
The variational wave function $(\ref{e})$ is the most natural choice
we can make. The wave function
$\psi_{010}(\rr_\e)\psi_{010}(\rr_\h)$ is simply the {\it free}
Hamiltonian $H_0$ ground state. Its presence in the trial function
$\phi(\rr_\e,\rr_\h)$ insures the validity of the perturbation
result to which the variational principle leads. Furthermore, in
relative coordinates, the perturbation function
$\phi_{\textrm{\scriptsize rel}}(r_{\e\h})$ exhibits the exciton
behavior, which is that of a hydrogen-like atom with a charge carrier
mass $\mu$. Because of this analogy, we expect the variational
parameter $\sigma$ to depend on the Bohr radius as $\ds\sigma\propto
a^{*-1}$. Notably, in the weak confinement regime, the exciton
ground state, which is a bound state, should mean that
$\ds\sigma\approx 2a^{*-1}$, so that $\phi_{\textrm{\scriptsize
rel}}(r_{\e\h})=\ee^{-\frac{r_\textrm{\tiny eh}}{a^*}}$ --- which
is, up to a normalization constant, the ground state wave function
of an hydrogen-like atom with its appropriate Bohr radius.

Let us make a trivial remark which will be the basic argument for
our later purpose: the spherical shape of the QD explicitly breaks
the translation invariance of the Coulomb interaction. This remark
on translation invariance and spherical symmetry breakdown, as a
whole, suggests the use of a Fourier transform formalism in relative
coordinates. Let $\mathcal F[f]$ stands for the Fourier transform of
the function $f$. Therefore, important matrix elements, such as the
square of the variational wave function norm, the diagonal matrix element
of the Coulomb electron-hole potential and the mean value of the
electron-hole {\it free} Hamiltonian $H_0$ in this variational
state, may be advantageously computed with Fourier transforms of
functions.

The square of the norm of $\phi$ is then given by
%\begin{widetext}
  \begin{eqnarray}
\lnd\rnd\langle\phi|\phi\rangle \nt
\\
\lnd=\rnd\int\!\dd^3\rr_\e\dd^3\rr_\h\psi^2_{010}(\rr_\e)\psi^2_{010}(\rr_\h)\phi^2_{\textrm{\scriptsize rel}}(\rr_{\e\h}) \nt
\\
\lnd=\rnd\int\!\frac{\dd^3\kk}{(2\pi)^3}\mathcal F\!\left[\phi_{\textrm{\scriptsize rel}}^2\right]\!\!(\kk)\mathcal F\!\left[\psi_{010}^2\right]\!\!(\kk)^2 \nt
\\
\lnd=\rnd\frac{-8}{R^2}\pr_\sigma\frac1\sigma\!\int\!\!\!\!\int_{\mathcal D}\!\frac{\dd x}x\frac{\dd y}y\sin^2(\pi x)\sin^2(\pi
y)\sinh(\sigma Rx)\ee^{-\sigma Ry}, \nt
\\
 \label{b}
  \end{eqnarray}
%\end{widetext}
where $\mathcal D=\left\{(x,y)\in\mathbb R^2/0\leq x\leq
y\leq1\right\}$, while the different Fourier transforms are
  \begin{eqnarray*}
\left\{
    \begin{array}{rcl}
\ds \mathcal F\!\left[\phi_{\textrm{\scriptsize rel}}^2\right]\!\!(\kk)\lnd=\rnd\ds -4\pi\pr_{\sigma}\frac1{\sigma^2+k^2},
\vs{.2cm}
\\
\ds \mathcal F\!\left[\psi_{010}^2\right]\!\!(\kk)\lnd=\rnd\ds \frac2{kR}\int_0^1\!\frac{\dd x}x\sin^2(\pi x)\sin(kRx).
    \end{array}
  \right.
  \end{eqnarray*}
Moreover, the diagonal matrix element of the electron-hole Coulomb potential is
%\begin{widetext}
  \begin{eqnarray}
\lnd\rnd\langle\phi|V_\C(r_{\e\h})|\phi\rangle \nt
\\
\lnd=\rnd-\frac{e^2}\kappa\!\int\!\frac{\dd^3\kk}{(2\pi)^3}\mathcal F\!\left[\varphi_{\textrm{\scriptsize rel}}\right]\!\!(\kk)\mathcal F\!\left[\psi_{010}^2\right]\!\!(\kk)^2\nt
\\
\lnd=\rnd\frac{-e^2}{\kappa R}\frac8\sigma\!\int\!\!\!\!\int_{\mathcal D}\!\frac{\dd x}x\frac{\dd y}y\sin^2(\pi x)\sin^2(\pi y)\sinh(\sigma Rx)\ee^{-\sigma Ry}, \nt
\\
 \label{c}
  \end{eqnarray}
%\end{widetext}
with $\ds\varphi_{\textrm{\scriptsize
rel}}(\rr_{\e\h})=\frac{|\phi_{\textrm{\scriptsize
rel}}(r_{\e\h})|^2}{r_{\e\h}}$. Lastly, the mean value
of the electron-hole Hamiltonian $H_0$ can be exactly evaluated as
  \begin{equation}
\frac{\langle\phi|H_0|\phi\rangle}{\langle\phi|\phi\rangle}=\frac{\pi^2}{2\mu
R^2}+\frac{\sigma^2}{8\mu}=\frac{\pi^2}{2\mu
R^2}+\frac{E^*}4\sigma'^2, \label{d}
  \end{equation}
where the dimensionless variational parameter $\sigma'$ is defined
by $\ds\sigma=\frac{\sigma'}{a^*}$ and the exciton Rydberg energy by
$\ds E^*=\frac1{2\mu a^{*2}}$.

As Eqs. (\ref{e}), (\ref{b}), (\ref{c}) and (\ref{d}) are obtained
without any further approximations --- except the ones upon which
the model was built ---, they have to be valid everywhere whether it
is, by construction, in the strong confinement regime or, by
extension, in the weak confinement regime. Indeed, the Coulomb
interaction itself should be always treated as a perturbation with
respect to the infinite potential well, although its energetic
contribution should not be maintained as a perturbation to the
exciton confinement energy. Thus, the trial function global form
(\ref{e}) will remain acceptable in the weak confinement regime, but
we will have to correct it by adding a further phase factor, which
will depends only on the center-of-mass coordinates.

We now study the behavior of these quantities and describe some of
their consequences in different ranges of QD radii for which $\sigma
R\ll1$ or $\sigma R\gg1$, (corresponding respectively to the strong
and the weak confinement regimes) and for which we can analytically
and explicitly compute numerical values.
\section{Strong confinement regime} \label{l}
In this regime, where $\sigma R\ll1$, the assumed form of the
variational wave function (\ref{e}) shall be used and it is
appropriate to perform Taylor expansions in Eqs. (\ref{b}) and
(\ref{c}) in the neighborhood of the dimensionless parameter $\sigma
R=0$. Because of the analyticity of the Taylor expansion of the
functions $\exp(x)$ and $\sinh(x)$, this expansion will remain valid
up to $\sigma R\lesssim1$. Thus, we obtain
  \begin{equation}
    \left\{\!\!
      \begin{array}{rcl}
\langle\phi|V_\C(r_{\e\h})|\phi\rangle\lnd=\rnd\ds-\frac{\ee^2}{\kappa R}\!
\left\{\!A-\sigma R+\frac B2\sigma^2R^2+\mathcal O(\sigma^3R^3)\!\right\}\!,
\vs{.2cm}
\\
\langle\phi|\phi\rangle\lnd=\rnd 1-B\sigma R+C\sigma^2R^2+\mathcal
O(\sigma^3R^3),
      \end{array}
    \right.
  \end{equation}
where the constants $A,B$ and $C$ are given in closed form by
  \begin{eqnarray*}
A\lnd=\rnd2-\frac{2\Si(2\pi)-\Si(4\pi)}{2\pi}\approx1.786,
\\
B\lnd=\rnd\frac23\left\{\frac{10}9-\frac2{3\pi^2}+\frac{2\Si(2\pi)-\Si(4\pi)}{8\pi^3}\!\right\}\!\approx0.699, \nt
\\
C\lnd=\rnd\frac13-\frac1{2\pi^2}\approx0.283. \nt
  \end{eqnarray*}
Here, $\Si(x)$ denotes the sine integral
  $$
\ds\Si(x)=\int_0^x\frac{\dd t}t\sin(t),~\forall x\in\mathbb R.
  $$
Consequently, we can deduce an expression of the mean value of the
total Hamiltonian $H$ in the strong confinement regime as an
expansion in powers of $\sigma'$, which is the dimensionless
variational parameter first introduced in Eq. (\ref{d})
  $$
\frac{\langle\phi|H|\phi\rangle}{\langle\phi|\phi\rangle}=
\frac{\pi^2}{2\mu R^2}-A\frac{e^2}{\kappa R}-2(AB-1)E^*\sigma'+\frac{E^*}4\sigma'^2+\dots
  $$
where the correction terms ``\dots'' go to zero at least as fast as
$\ds\propto\frac R{a^*}$. The variational parameter $\sigma'$ is now
determined by minimizing the expectation value of the energy. Thus,
we find $\sigma'_0=4(AB-1)\approx0.996$ and the corresponding energy value is
  \begin{eqnarray}
E^{\textrm{\scriptsize strong}}_{\e\h}\lnd=\rnd\frac{\pi^2}{2\mu R^2}-A\frac{e^2}{\kappa R}-4(AB-1)^2E^* \nt
\\
\lnd\approx\rnd\frac{\pi^2}{2\mu R^2}-1.786\frac{e^2}{\kappa R}-0.248E^*.
  \end{eqnarray}
In the strong confinement regime, this formula has been already
analytically obtained \cite{Kayanuma_1988, Schmidt_1986} with trial
functions showing the same global form as the previously presented
one (\ref{e}) but with an interactive part chosen, instead of
$\phi_{\textrm{\scriptsize rel}}(\rr_{\e\h})$, equal to
  $$
\widetilde\phi_{\textrm{\scriptsize
rel}}(\rr_{\e\h})=1-\frac\sigma2r_{\e\h}.
  $$
This choice corresponds
in fact to the two first terms of the Taylor expansion of Eq.
(\ref{e}) in the neighborhood of $\frac\sigma2r_{\e\h}\leq\sigma
R\ll1$. In the meantime, we have succeeded in finding an approximate
value of its upper bound for $R\lesssim a^*$. This bound, as we
already mentioned in Subsec. {\bf\ref{h}}, can be in turn
numerically extend to the commonly accepted region of validity of QD
radii, {\it i.e.} $R\lesssim 2a^*$.
\section{Weak confinement regime} \label{m}
As previously explained, in this regime $\sigma R\gg1$, we can
retain the global form of the trial function used in the strong
confinement regime because of the presence of the confinement
potential. But, as said before, the physical situation requires a
modification, before applying the variational procedure.
\subsection{Considerations on the trial function}
Here, the electron-hole pair states consist in binding exciton
states, which globally act like quasi-particles of
total mass $M$. As matter of fact, in a translation invariant space
region, the quasi-particle point of view emphasizes the possibility
for the exciton to show a global translational motion in terms of
its center-of-mass coordinates. In the weak confinement regime, the
exciton typical size $\propto a^*$ is sufficiently smaller than the
QD typical radius $R$. So, we can reasonably assume that excitons
should exhibit a quasi-particle behavior and possess a global
translational motion. In order to account for a partial restoration
of translation invariance, in the weak confinement regime, we have
to introduce a center-of-mass coordinates plane wave, because the
physical behavior of the confined exciton should show a reminiscence
of the free exciton one, {\it i.e.} the wave function of the ground
state of the exciton shall be considered as the wave function of
quasi-particle describing the electron-hole pair. Thus, it should
have the form
    \begin{equation}
\psi(\rr_\e,\rr_\h)=\psi_{010}(\rr_\e)\psi_{010}(\rr_\h)
\phi_{\textrm{\scriptsize rel}}(\rr_{\e\h})\phi_\e(\rr_\e)
\phi_\h(\rr_\h), \label{i}
  \end{equation}
where $\phi_{\textrm{\scriptsize e,h}}(\rr_{\e,\h})=\e^{i\frac\pi
R\textrm{\scriptsize$\sigmag$}_{\!\textrm{\tiny
e,h}}\cdot\rr_{\textrm{\tiny e,h}}}$. The vectors $\sigmag_{\e,\h}$
are dimensionless quantities which facilitate the computation in
practice but are not directly physically interpretable. In fact, in
order to understand properly the meaning of these parameters, we
have to define the wave numbers vectors respectively
in the center-of-mass and in the relative coordinates as
  $$
\left\{
  \begin{array}{rcl}
\sigmag_\G\lnd=\rnd\sigmag_\e+\sigmag_\h,
\vs{.1cm}
\\
\sigmag_{\e\h}\lnd=\rnd\ds\frac{m_\h^*\sigmag_\e-m_\e^*\sigmag_h}M.
  \end{array}
\right.\!\!
  $$
Since $\sigmag_\e\cdot\rr_\e+\sigmag_\h\cdot\rr_\h=
\sigmag_\G\cdot\rr_\G+\sigmag_{\e\h}\cdot\rr_{\e\h}$, the functions
$\phi_{\e,\h}(\rr_{\e,\h})$ should contribute to
the exciton total energy by an additional kinetic term,
corresponding to the fundamental energy of a plane wave in a space
region of size $R$ in the center-of-mass coordinates $\rr_\G$,
$\propto\ds\frac{|\sigmag_\e|^2}{m_\e^*}+\frac{|\sigmag_\h|^2}{m_\h^*}=
\frac{|\sigmag_\G|^2}M+\frac{|\sigmag_{\e\h}|^2}\mu$, which is
assumed to be of the form $\ds\frac{\pi^2}{2MR^2}$. This constraint
is consistent with the conditions
  \begin{equation*}
\left\{
  \begin{array}{rcl}
|\sigmag_\G|^2\lnd=\rnd1
\vs{.1cm}
\\
\sigmag_{\e\h}\lnd=\rnd\ds\zerog
  \end{array}
\right.\Longrightarrow~\left\{
  \begin{array}{rcl}
|\sigmag_\ee|^2\lnd=\rnd\ds\frac1{(1+\lambda)^2}
\vs{.1cm}
\\
|\sigmag_\h|^2\lnd=\rnd\ds\frac{\lambda^2}{(1+\lambda)^2}
  \end{array}
\right.\!
  \end{equation*}
where $\ds\lambda=\frac{m_\h^*}{m_\e^*}$. The previous expressions reinforce the cogency of the trial wave
function $\psi(\rr_\e,\rr_\h)$. Taking $\sigmag_{\e\h}=\zerog$ does
not provide an extra kinetic energy term to the total exciton energy
in the relative coordinates, except the ones due to the electron and
hole confinement. But taking $|\sigmag_\G|^2=1$ adds to the
electron-hole pair total energy the correct energetic contribution
in the center-of-mass coordinates, corresponding to the fundamental
energy of a confined particle of mass $M$.
\subsection{Variational principle}
The trial function (\ref{i}) leaves unchanged the exciton density
of probability or the Coulomb potential matrix element
  \begin{equation*}
\langle\psi|\psi\rangle=\langle\phi|\phi\rangle\textrm{~~and~~}
\langle\psi|V_\C(r_{\e\h})|\psi\rangle=
\langle\phi|V_\C(r_{\e\h})|\phi\rangle,
  \end{equation*}
whereas the free Hamiltonian matrix element gets further contributions, {\it i.e.}
  \begin{eqnarray*}
\frac{\langle\psi|H_0|\psi\rangle}{\langle\psi|\psi\rangle}\lnd=\rnd
\frac{\pi^2}{2\mu R^2}+\frac{\pi^2}{2R^2}\!\left\{\frac{|\sigmag_\e|^2}{m_\e^*}+
\frac{|\sigmag_\h|^2}{m_\h^*}\right\}\!+\frac{E^*}4\sigma'^2
\\
\lnd=\rnd\frac{\pi^2}{2\mu R^2}+\frac{\pi^2}{2MR^2}+\frac{E^*}4\sigma'^2.
  \end{eqnarray*}

In order to evaluate the mean value of the Coulomb potential in the
quantum state defined by the wave function $\psi$, we have to
compute the following double integral in the weak confinement regime
for $\ds\sigma R\gtrsim2\pi$.  Since for such QD radii, the
convergence of all occurring series is insured, we can write
\begin{widetext}
  \begin{eqnarray}
\lnd\rnd\int\!\!\!\!\int_{\mathcal D}\!\frac{\dd x}x\frac{\dd y}y\sin^2(\pi x)\sin^2(\pi y)\sinh(\sigma Rx)\ee^{-\sigma Ry} \nt
\\
\lnd=\rnd\int\!\!\!\!\int_{0\leq x\leq y\leq\sigma R}\!\frac{\dd x}x\frac{\dd y}y\sin^2\!\left(\frac\pi{\sigma R}x\right)\!\sin^2\!\left(\frac\pi{\sigma R}y\right)\!\sinh(x)\ee^{-y} \nt
\\
\lnd=\rnd\frac{\pi^2}{2\sigma^2R^2}\sum_{k\geq0}\frac1{k+1}\!\left(\!-\frac{4\pi^2}{\sigma^2 R^2}\!\right)^{\!\!k}\!\int_0^{\sigma R}\!\frac{\dd y}y\sin^2\!\left(\frac\pi{\sigma R}y\right)\!\!\left\{\ee^{-2y}\sum_{n=0}^{2k+1}\frac{y^n}{n!}-\sum_{n=0}^{2k+1}\frac{(-y)^n}{n!}\right\} \nt
\\
\lnd=\rnd\frac{2\pi^2}{\sigma R}\sum_{n\geq0}\frac{(-4\pi^2)^n}{(2(n+1))!}
\!~_2\F_1\!\left(n+1,1;n+2;-\frac{4\pi^2}{\sigma^2R^2}\right)\! \nt
\\
\lnd\rnd~~~~\times\!\left\{\int_0^1\!\frac{\dd y}y\sin^2(\pi
y)y^{2n+1}\cosh(\sigma Ry)-\frac{2n+1}{\sigma R}\int_0^1\!\dd
y\!~y^{2n}\sinh(\sigma Ry)\!\right\}\!,\label{p}
  \end{eqnarray}
\end{widetext}
where
$\ds_2\F_1(a,b;c;z)=\sum_{n\geq0}\frac{(a)_n(b)_n}{(c)_n}\frac{z^n}{n!}$
is the usual Gauss hypergeometric function. In Eq. (\ref{p}),
we take terms up to the fourth order in the variable
$\ds\frac\pi{\sigma R}$ in the neighborhood of $0$. Therefore, for
the terms in Eq. (\ref{p}) whose integrand is not exponentially
decreasing for $y\rightarrow\sigma R\gg1$, we get
\begin{widetext}
  \begin{eqnarray}
\lnd\rnd-\frac{\pi^2}{2\sigma R}\int_0^{\sigma R}\!\frac{\dd y}y\sin^2\!\left(\frac\pi{\sigma R}y\right)\sum_{k\geq0}\frac1{k+1}\!\left(\!-\frac{4\pi^2}{\sigma^2 R^2}\!\right)^{\!\!k}\sum_{n=0}^{2k+1}\frac{(-y)^n}{n!} \nt
\\
\lnd=\rnd A'-\frac{B'}{\sigma R}+C'\frac{\pi^2}{\sigma^2R^2}-D'\frac{\pi^3}{\sigma^3R^3}+E'\frac{\pi^4}{\sigma^4R^4}+\mathcal O\!\left(\frac{\pi^5}{\sigma^5R^5}\right)\!; \label{f}
  \end{eqnarray}
\end{widetext}
with
  \begin{eqnarray*}
A'\lnd=\rnd\frac\pi2\!\left\{\Si(2\pi)-\frac{\Si(4\pi)}2\right\}\!\approx1.056,~B'=0,~
\\
C'\lnd=\rnd-\frac43A',~D'=\frac\pi4\textrm{~~and~~}E'=\frac45\!\left(\!A'-\frac18\right)\!.
  \end{eqnarray*}
Furthermore, for terms in Eq. (\ref{p}) whose integrand is
exponentially decreasing for $y\rightarrow\sigma R\gg1$, we apply
Laplace method ({\it cf.} Appendix {\bf\ref{q}}) to get an answer in
the limit of $\sigma R\rightarrow\infty$. Hence,

\begin{widetext}
  \begin{eqnarray}
\pi^2\lnd\ds\sum_{n\geq0}\rnd\frac{(-4\pi^2)^n}{(2(n+1))!}
\!~_2\F_1\!\left(n+1,1;n+2;-\frac{4\pi^2}{\sigma^2R^2}\right)\!\!\int_0^1\!\frac{\dd y}y\ee^{-2\sigma Ry}\sin^2(\pi y)\!\left\{y^{2n+1}+\frac{2n+1}{\sigma R}y^{2n}\right\}\! \nt
\\
\lnd=\rnd\frac{\pi^3}{\sigma^3R^3}\!\!\left\{\frac\pi4+\mathcal
O\!\left(\frac{\pi^2}{\sigma^2R^2}\right)\!\right\}\!.
  \end{eqnarray}
\end{widetext}
In the QD radii range, where the third order actually contributes to
Eq. (\ref{f}), {\it i.e.} for radii near the lower bound of the
possible radii in the weak confinement regime $\ds\sigma
R\approx2\pi$ --- which, as we will see later, corresponds to a size
$R\approx\pi a^*$ ---, the third and fourth order terms contribute
to the exciton energy with terms of the same order of magnitude. But
for radii larger than $4a^*$ \cite{Kayanuma_1988}, the third and
also the fourth order terms are in fact irrelevant. So, we can stop
the expansion in Eq. (\ref{f}) at the second order term. Thus, the Coulomb interaction diagonal matrix element and the square of the norm become
\begin{widetext}
  \begin{equation}
    \left\{
      \begin{array}{rcl}
\langle\psi|V_\C(r_{\e\h})|\psi\rangle\lnd=
\rnd\ds-\frac{e^2}{\kappa R}\frac{8A'}{\sigma^2R^2}\left\{1-\frac43\frac{\pi^2}{\sigma^2R^2}+\frac{D''}{A'}\frac{\pi^3}{\sigma^3R^3}+\mathcal O\!\left(\frac{\pi^4}{\sigma^4R^4}\right)\right\}\!,
\vs{.2cm}
\\
\langle\psi|\psi\rangle\lnd=\rnd\ds\frac{16A'}{\sigma^3R^3}
\left\{1-\frac83\frac{\pi^2}{\sigma^2R^2}+\frac52\frac{D''}{A'}\frac{\pi^3}{\sigma^3R^3}+\mathcal O\!\left(\frac{\pi^4}{\sigma^4R^4}\right)\right\}\!,
      \end{array}
    \right.
  \end{equation}
\end{widetext}
where by definition $D''=\ds\frac{E'}2\approx0.372$. Finally, we
obtain in this regime the following $\sigma'$-expansion of the
variational energy
%\begin{widetext}
  \begin{eqnarray*}
\frac{\langle\psi|H|\psi\rangle}{\langle\psi|\psi\rangle}\lnd=\rnd
\frac{\pi^2}{2\mu R^2}+\frac{\pi^2}{2MR^2}+\frac{E^*}4\sigma'^2-E^*\sigma'-\frac23\frac{\pi^2}{\mu R^2}\frac1{\sigma'}
\\
\lnd\rnd~~~~+\frac34\frac{D''}{A'}\pi\frac{\pi^2}{\mu R^2}\frac1{\sigma'^2}\frac{a^*}R+\dots
  \end{eqnarray*}
%\end{widetext}
Applying now the variational principle, we get the value of the dimensionless variational parameter
$\sigma'$
  $$
\ds
\sigma'_0=2-\frac23\pi^2\!\left(\frac{a^*}R\right)^{\!\!2}\!+\frac34\frac{D''}{A'}\pi^3\!\left(\frac{a^*}R\right)^{\!\!3}\!\!.
  $$
Because the second and the third terms in the previous expression do not contribute to the
total electron-hole pair energy, we conclude that
$\sigma'_0\approx2$ and
  \begin{eqnarray}
E^{\textrm{\scriptsize weak}}_{\e\h}\lnd=\rnd-E^*+\frac{\pi^2}{6\mu R^2}+\frac{\pi^2}{2M R^2}+\delta\frac{\pi^2}{\mu R^2}\frac{a^*}R \nt
\\
\lnd=\rnd-E^*+\frac{\pi^2}{6\mu R^2}+\frac{\pi^2}{2M(R-\eta(\lambda)a^*)^2}, \label{g}
  \end{eqnarray}
from which we are able to extract an explicit expression for the
Kayanuma-function as
$\ds\eta(\lambda)=\delta\frac{(1+\lambda)^2}\lambda$ with $\delta=\ds\frac3{16}\frac{D''}{A'}\pi=\frac{3\pi}{40}\!\left(1-\frac1{8A'}\right)\!\approx0.208$ and $\ds\lambda=\frac{m_\h^*}{m_\e^*}$. As we have already specified, let us observe that
$\sigma'_0\approx2$ implies that $\ds R\gtrsim\pi a^*$, which is
\emph{surprisingly} consistent with the usually accepted numerical
region of validity. Moreover, we see that the interaction part
$\phi_{\textrm{\scriptsize rel}}(\rr_{\e\h})$ of the wave function
coincides with the ground state wave function of a reduced mass
$\mu$ hydrogen-like atom up to a normalization factor and
contributes to the excitonic total energy with the leading binding
fundamental energetic term $-E^*$, as we expect.

The $\eta$-function was first introduced phenomenologically by Eq.
(28) in \cite{Kayanuma_1988} to get a better fit between numerical
and empirical results; but no analytic derivation of it is available
so far. First, here, we manage to give an expression for this
function, which satisfies the electron-hole exchange symmetry:
$\lambda\longrightarrow\lambda^{-1}$. Second, as was already
depicted by Kayanuma, the QD size {\it renormalization} term
$\eta(\lambda)a^*$ is interpretable as the so-called dead layer
\cite{Hopfield_1963}: this is the physical reminiscence of the fact
that, although it could be successfully described as a
quasi-particle, the exciton is not actually itself an indivisible
particle. In fact, its center-of-mass motion, the one on which the
quantization is really performed, could not reach the infinite
potential well surface unless the exciton undergoes a strong
deformation in the relative motion of the electron and the hole.
This implies that the picture of a point-like exciton is no longer
appropriate in this region of space. The most convenient way to
study this exciton consists in thinking of it as a rigid sphere of
radius $\propto a^*$, where the proportional factor must not exceed
$\frac32$ too much, because $l^*=\frac32a^*$ is the mean value of
the relative distance between the electron and the hole in the
non-confined exciton ground state. We have to get the largest
possible radius $l^*$ for the sphere picture, when
$\lambda\rightarrow\infty$, {\it i.e.} in the limit of infinite hole
mass because the hole stays motionless at the center-of-mass of the
electron-hole system. Furthermore, the smallest possible radius must
be obtained in the symmetrical case $\lambda=1$ and must be held at
$\frac12l^*=\frac34a^*$, which imposes $\eta(1)\approx0.75$. This
matches quite well with both experimental and theoretical results as
shown in Tab. \ref{table_1}. Finally, regarding the different values
of $\eta(\lambda)$ around $\frac32$ at $\lambda=5$, we can safely
conclude that the picture of infinite hole mass becomes valid as
soon as $\lambda\approx5$.
\begin{table}[h]
\caption{\label{table_1}Comparison between numerical and analytical values for the
$\eta$-function.}
  \begin{ruledtabular}
    \begin{center}
      \begin{tabular}{cccc}
$\lambda$ & 1 & 3 & 5
\\
\hline
$\eta_{\textrm{\scriptsize num}}(\lambda)$ & 0.73 & 1.1 & 1.4
\\
$\eta_{\textrm{\scriptsize theo}}(\lambda)$ & 0.83 & 1.1 & 1.5
\\
relative error & 14\% & $<$1\% & 7\%
\\
      \end{tabular}
    \end{center}
  \end{ruledtabular}
\end{table}

In Tab. \ref{table_1}, we compare numerical values computed from
numerical simulations taken by this function $\eta(\lambda)$ for
$\lambda=1,3,5$ to those theoretically predicted by the previous
relation and note that there exists a reasonable agreement between
both results. We even succeed to enlarge the region, in which the
weak confinement regime is satisfactory.
\section{A pseudo-potential-like method} \label{n}
The problem, that we still have to deal with, consists in finding a
physical way to substract off the term reminiscent
of the kinetic energy term in Eq. (\ref{g}). As matter of fact,
because of the presence of the reduced mass, it is interpretable as
a kinetic energy term in the relative coordinates. However, this
type of exciton kinetic energy has to be already contained in the
Rydberg energy term, because of the validity of the Virial theorem
in the relative coordinates. Besides, in this assumption, the higher
order contributions to the electron-hole energy must be interpreted
as higher order contributions to the kinetic energy of the exciton,
viewed as a quasi-particle of mass $M$, which physically justifies
the intuitive idea that for very large radius $R$ only the
quasi-particle point of view should be responsible for the exciton
kinetic energy.
\subsection{Weak confinement regime}
In analogy with the pseudo-potential method for metals, we propose
to introduce an additional potential term $W$ in the
Hamiltonian $H$, which shall depend only on the electron-hole
relative distance $r_{\e\h}$, so that it contributes to the second order of the
exciton total energy but not to the third order --- {\it i.e.} the
one which allows us to determine the expectation value of the
$\eta$-function. Furthermore, we assume that $W$ must
  \begin{itemize}
\item be attractive at distances in the range of $a^*$ in order to
promote excitonic state with typical size around his Bohr radius;
\item be repulsive at short distances in order to penalize excitonic
state with small size and to remind that the exciton typical size
should not be smaller that the typical size of the surrounding
lattice spacing;
\item be exponentially small for large distances, in order to not perturb the long range Coulomb interaction.
  \end{itemize}
Consequently a natural choice of such potential should be of the
form
  \begin{equation}
W(\rr_{\e\h})=-\frac{32\pi^2}9E^*\frac{r_{\e\h}^2}{R^2}\ee^{-2\frac{r_{\textrm{\tiny
eh}}}{a^*}}.
\label{j}
  \end{equation}
Inspection shows that it satisfies all of the previous constraints;
and, in the weak confinement regime, it exactly provides us what we
ask for
  \begin{equation*}
\frac{\langle\psi|W(\rr_{\e\h})|\psi\rangle}{\langle\psi|\psi\rangle}=
-\frac{\pi^2}{6\mu R^2}\!\left\{1+\mathcal O\!\left(\frac{a^{*2}}{R^2}\right)\!\right\}\!.
  \end{equation*}
\textbf{Remark}. The potential amplitude in Eq. (\ref{j}) is fixed {\it
ad hoc} in order to get the previous correct kinetic energy
contribution. Consequently, the pseudo-potential form we choose
seems to be arbitrary. However, the form of $W(r_{\e\h})\propto
r_{\e\h}^2\ee^{-2\frac{r_\textrm{\tiny eh}}{a^*}}$ is the only one
which straightforwardly first contributes to the second order term
in both weak and strong confinement regimes and which does not
change the exciton energy behavior to zeroth and first order terms.
\subsection{Strong confinement regime}
As we have introduced a new term in the exciton total Hamiltonian
$H$, we have to check its consequences in the strong confinement
regime. The most important one is the significant decrease of the
expected value of the exciton energy because of the exciton total
Hamiltonian redefinition: $H'=H+W$. In fact, the pseudo-potential
$W$ contributes to the second order as
  \begin{eqnarray*}
\frac{\langle\phi|W(\rr_{\e\h})|\phi\rangle}{\langle\phi|\phi\rangle}\lnd=\rnd
-\frac{64\pi^2}9CE^*\!\left\{1+\mathcal O\!\left(\frac R{a^*}\right)\!\right\}
\\
\lnd\approx\rnd-19.9E^*+\dots
  \end{eqnarray*}
This expression is deduced by a reasoning similar to that of Sec.
{\bf \ref{l}}. Therefore, we must keep in mind that it is only valid when
$2R\lesssim a^*$ because of the pseudo-potential exponential
dependence, which changes the validity conditions of the Taylor
expansion we made.
  \begin{figure}[h]
\caption{Behavior of the excitonic ground state energy as a function
of the QD radius computed for a confining infinite potential well
with (--$\!~$--) or without (---) the presence of the
pseudo-potential $W$ and for a confining finite potential step of
height $V_0\approx1$eV (--$~\!\cdot\!~$--) \cite{Thoai_1990} and compared to
experimental results for $CdS$ \cite{Weller_1986} microcrystallites
with material parameters: $\varepsilon=11.6$, $m_\e^*=0.235m_\e$,
$m_\h^*=1.35m_\e$, $E_\g=2.583$eV, $E^*=27$meV and $a^*=30.1$\AA, where $m_\e$ is the electron bare mass.}
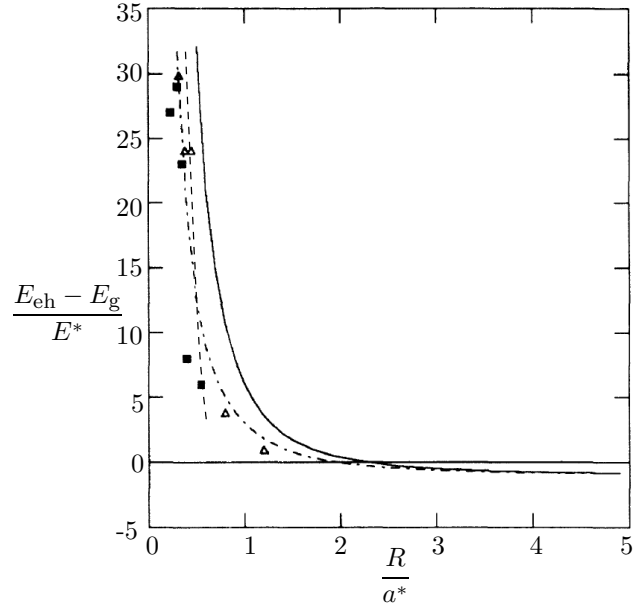
\label{figure_1}
    \begin{center}
\input{fig1.pstex_t}
    \end{center}
  \end{figure}

As we can see from Fig. \ref{figure_1},
the excitonic energy computed in the presence of the
pseudo-potential gets a {\it significant} better fit to experimental
results, in the validity domain $2R\lesssim a^*$, than the one
calculated without this tool. Nevertheless, the divergence for very
small QD size still holds as a consequence of the infinite potential
well assumption. In order to extend the validity domain upper bound,
it would be efficient to carry on the energy expansion to one or
maybe to two further orders. But, at this point, the computation
become quite involved and we cannot insure the success or even the
relevance of this approach.
\section{Conclusion}
By using an improved EMA model, we are able to obtain well known
results in the strong confinement regime, to correctly apprehend the
weak confinement regime by adding a pseudo-potential term to the
total Hamiltonian describing the electron-hole dynamics and to find
an nice analytical expression for the phenomenological function
$\eta$ of Kayanuma.

As we have shown, the introduction of the pseudo-potential $W$
allows to reduce in part the overestimation due to the infinite
potential well observed in \cite{Thoai_1990}, particularly near the
upper boundary $2R\approx a^*$, even if it still does not accurately
approximate the total energy for much smaller QDs. For such QD
sizes, the best approach would probably require a finite confinement
potential. Thus, a future research work might focus on applying this
pseudo-potential method, in the presence of a confining finite
potential step, in order to obtain a hopefully
better behavior for excitons in very small QDs.

\appendix

\section{The Laplace method} \label{q}
The purpose of this
method is to study asymptotic behavior of integrals like $\ds
I(t)\!=\!\!\!\int_0^a\!\!\dd x\!~g(x)\ee^{th(x)}$ and rigorously
determine a mathematical equivalent for such quantities when
$t\rightarrow\infty$.
\\
One can prove the following theorem \cite{Gourdon_1994}: Let $a>0$
and $g,h:[0,a]\longrightarrow\mathbb R$ two continuous mappings such that
  \begin{itemize}
\item[{\bf i.}] $\ds\int_0^a\!\dd x\!~|g(x)|\ee^{h(x)}<\infty$;
\item[{\bf ii.}] $\exists\delta_0>0~/~\forall\delta\in[0,\delta_0]$ and $\forall x\in[\delta,a[,~h(x)\leq h(\delta)$;
\item[{\bf iii.}] $g(x)\stackrel{\sim}{x0^+}
Ax^\alpha\textrm{~~and~~}h(x)=b-cx^{\beta}+o(x^\beta)$, where $\alpha>-1$ and $c,\beta>0$.
  \end{itemize}
Then, $\ds I(t)\stackrel{\sim}{t\infty}\frac
A\beta\Gamma\!\left(\frac{\alpha+1}\beta\right)\!\ee^{bt}(ct)^{-\frac{\alpha+1}\beta}$.
\end{document}